\documentclass[]{spie}  %>>> use for US letter paper

\usepackage{subfigure}
\usepackage{amsmath,amsfonts,amssymb}
\usepackage[left=0.5in,right=0.5in,top=0.5in,bottom=0.5in]{geometry}
\usepackage{graphicx}
\usepackage{booktabs}
\usepackage[linesnumbered,ruled,vlined]{algorithm2e}
\usepackage{multirow}
\usepackage{floatrow}
\usepackage[colorlinks=true, allcolors=blue]{hyperref}
\usepackage[font=small]{caption}
\graphicspath{ {./figure/} }

\newfloatcommand{capbtabbox}{table}[][\FBwidth]
\SetKwInput{KwInput}{Input}                % Set the Input
\SetKwInput{KwOutput}{Output}  
\title{Estimating Task-based Performance Bounds for Accelerated MRI Image Reconstruction Methods by Use of Learned-Ideal Observers}

\author[a]{Kaiyan Li}
\author[b]{Prabhat Kc}
\author[a,c,d]{Hua Li}
\author[e]{Kyle J. Myers}
\author[a]{Mark A. Anastasio}
\author[b]{Rongping Zeng}

\affil[a]{Department of Bioengineering, University of Illinois Urbana-Champaign, Urbana, IL, USA}
\affil[b]{Center for Devices and Radiological Health, Food and Drug Administration, Silver Spring, MD, USA}
\affil[c]{Department of Radiation Oncology, Washington University in St. Louis, Saint Louis, MO, USA}
\affil[d]{Cancer Center at Illinois, Urbana, IL, USA}
\affil[e]{Puente Solutions LLC, Phoenix, AZ, USA}

\authorinfo{
Send correspondence to Mark A. Anastasio \& Rongping Zeng. E-mail: \{maa@illinois.edu \& Rongping.Zeng@fda.hhs.gov\}.}
\begin{document} 
\maketitle

%\vspace{-0.3cm}   
\begin{abstract}

Medical imaging systems are commonly assessed
and optimized by the use of objective measures of
image quality (IQ). The performance of the ideal observer (IO) acting on imaging measurements
has long been advocated as a figure-of-merit to guide the optimization of imaging systems.
For computed imaging systems, the performance of the IO acting on imaging measurements also sets an upper bound on task-performance that no image reconstruction method can transcend.
As such, estimation of IO performance can provide valuable guidance when designing under-sampled data-acquisition techniques by
enabling the identification of designs that will not permit the reconstruction of diagnostically inappropriate images for a specified task - no matter how advanced the reconstruction method is or how plausible the reconstructed images appear. The need for such
analysis is urgent because of the substantial increase of medical device submissions on deep learning-based image reconstruction methods and the fact that they may produce clean images disguising the potential loss of diagnostic information when data is aggressively under-sampled.
Recently, convolutional neural network (CNN) approximated IOs (CNN-IOs) was investigated for estimating the performance of data space IOs to establish task-based performance bounds for image reconstruction, under an X-ray computed tomographic (CT) context.
%However, the application of such data space CNN-IO analysis to more realistic multi-coil magnetic resonance imaging (MRI) systems remains unexplored.
In this work, the application of such data space CNN-IO analysis to multi-coil magnetic resonance imaging (MRI) systems has been explored. 
%estimates of the data space IO were computed with consideration of 
This study utilized stylized multi-coil sensitivity encoding (SENSE) MRI systems and deep-generated stochastic brain models to demonstrate the approach. Signal-known-statistically and background-known-statistically (SKS/BKS) binary signal detection tasks were selected to study the impact of different acceleration factors on the data space IO performance.
\end{abstract}

%\vspace{-0.1cm}
\section{Introduction}
\label{sec:purpose}  % \label{} allows reference to this section
%In medical imaging, images are often acquired for specific purposes. 
%\textcolor{blue}{Objective measures of image quality (IQ) can be employed to quantify the data diagnostic value for the specified task.}
Objective measures of image quality (IQ) that quantify the ability of an observer to perform a specific task are widely employed in
the field of medical imaging \cite{barrett2013foundations,kupinski2003ideal, li2021assessing,zhou2019approximating,li2022task, li2022impact}.
For the purpose of optimizing imaging system designs, objective IQ measures based on the performance of 
the Bayesian Ideal Observer (IO) acting on imaging measurements with consideration of different diagnostic tasks have been advocated \cite{barrett2013foundations,kupinski2003ideal}.
In this way, the amount of task-specific information in the imaging measurements can be maximumly used.
Moreover, for computed imaging systems, the performance of the IO acting on imaging measurements sets an upper bound on task-performance that no image reconstruction method can improve upon.
The ability to compute performance bounds for computed imaging systems is now more important than ever for evaluating deep learning-based image reconstruction methods (DLIRMs).
A variety of DLIRMs are being actively developed, primarily for applications of under-sampled data acquisitions in which conventional image
reconstruction methods are not fully effective. For example, a possible use case for DLIRMs is image
reconstruction from highly incomplete and noisy tomographic measurements.
In some cases, DLIRMs can yield visually plausible images that possess encouraging scores as measured by
physical, non-task-based, metrics such as MSE, PSNR, and SSIM.
However, these metrics may disguise the situations in
which incomplete and noisy tomographic measurement data will not permit the reconstruction of a \emph{diagnostically useful} image, no matter how advanced the DLIRM is or plausible the reconstructed images appear.
This can occur when features of the object that are important for performing the diagnostic task are not sufficiently presented in the measurement data.
A natural way to identify such situations is to estimate the
performance of the IO acting on the measurement data as a performance bound for any image reconstruction methods.

Recently, convolutional neural network (CNN) approximated IOs (CNN-IOs) have been investigated for estimating the performance of data space IOs to establish task-based performance bounds for image reconstruction~\cite{li2024application}. 
The effectiveness of the data space CNN-IO has been demonstrated in a relatively simple simulation study relevant to the X-ray computed tomographic (CT) context.
The application of such data space CNN-IO analysis to advanced multi-coil magnetic resonance imaging (MRI) systems and more realistic object variability remains unexplored. Considering the active development of DLIRMs for multi-coil accelerated MRI acquisitions, there exists an urgent need to explore the applications of data space CNN-IO in MRI.

In this work, the application of data space CNN-IOs was explored in a  realistically simulated MRI imaging context.
The study simulates a multi-coil sensitivity encoding (SENSE) parallel MRI system.
To characterize more realistic object variability, an advanced diffusion model was trained with a large set of high-quality MRI brain images to establish the stochastic object model (SOM) of the human brain.
In addition, signal-known-statistically and background-known-statistically (SKS/BKS)  binary signal detection tasks with random signal locations were considered to study the impact of acceleration factors on task-based performance bounds for image reconstruction.

\section{Background}
\label{sec:background}
%%\vspace{-0.1cm}
\subsection{Formulation of binary signal detection tasks}
%\vspace{-0.1cm}
 A continuous-to-discrete (C-D) description of a linear imaging system \cite{barrett2013foundations} is considered as $\mathbf{g}=\mathcal{H}{f(\mathbf{r})+\mathbf{n}}$,
where $\mathbf{g}\in\mathbb{R}^{N\times 1}$ is the measured image vector, 
$f(\mathbf{r})$ denotes the object function that is dependent on the coordinate $\mathbf{r}\in\mathbb{R}^{k\times 1}$, $k \ge 2$,
$\mathcal{H}$ denotes a linear imaging operator that maps $\mathbb{L}_{2}(\mathbb{R}^{k})$ 
to $\mathbb{R}^{N\times 1}$, 
and $\mathbf{n}\in\mathbb{R}^{N\times 1}$ denotes the measurement noise. 
When its spatial dependence is not important to highlight, $f(\mathbf{r})$ will be denoted as $\mathbf{f}$. 
%\vspace{-0.1cm}

A  binary signal detection task requires an observer 
to classify the measured image data $\mathbf{g}$ as satisfying either a signal-present hypothesis $H_1$ or a signal-absent hypothesis $H_0$.
These two hypotheses can be described as:
\vspace{-0.1cm}
\begin{subequations}
	\label{eq:hypo}
	\begin{equation}
	H_0:\mathbf{g}=\mathcal{H}\mathbf{f_b}+\mathbf{n}=\mathbf{b+n}, 
	\end{equation}
	\begin{equation}
	H_1:\mathbf{g}=\mathcal{H}\mathbf{(f_b+f_s)}+\mathbf{n}=\mathbf{b+s+n},   
	\end{equation}   
\end{subequations}
\noindent where $\mathbf{f_s}$ and $\mathbf{f_b}$ denote the signal and background object,
respectively, 
and $\mathbf{s}\equiv\mathcal{H}\mathbf{f_s}$ and $\mathbf{b}\equiv\mathcal{H}\mathbf{f_b}$ denote the measured signal and background image data. 
To perform this task, a deterministic observer
computes a test statistic that maps the measured image $\mathbf{g}$ to a real-valued scalar variable that is compared to a
predetermined threshold $\tau$ to determine if $\mathbf{g}$ satisfies $H_0$ or $H_1$. 
By varying the threshold $\tau$, a ROC curve can be formed to quantify the trade-off between the false-positive fraction (FPF) and the true-positive fraction (TPF) \cite{barrett2013foundations}. The area under the ROC curve (AUC) can be subsequently calculated as a figure-of-merit (FOM) for signal detection performance.
%\vspace{-0.2cm}
\subsection{The data space ideal observer (IO) and the CNN-approximated data space IO}
\label{ssec:IO}
%\vspace{-0.1cm}
The Bayesian Ideal Observer (IO) sets an upper limit of observer performance for signal detection tasks 
and has been advocated for use in optimizing medical imaging systems and data-acquisition designs~\cite{barrett2013foundations}.
The IO test statistic $t_{\text{IO}}(\textbf{g})$ is any monotonic transformation of the likelihood ratio $\Lambda_{\text{LR}}(\mathbf{g})=\frac{p(\mathbf{g}|H_1)}{p(\mathbf{g}|H_0)}$,
%
% \begin{equation}
% \label{eq:LR}
% \Lambda_{\text{LR}}(\mathbf{g})=\frac{p(\mathbf{g}|H_1)}{p(\mathbf{g}|H_0)},
% \end{equation}
%
\noindent where $p(\mathbf{g}|H_1)$ and $p(\mathbf{g}|H_0)$ are the conditional probability density functions that describe the measured data $\mathbf{g}$ under the hypotheses $H_1$ and $H_0$, respectively.
In general, the likelihood ratio $\Lambda_{\text{LR}}(\mathbf{g})$ is often analytically intractable.
In this study, a supervised learning-based method is employed to approximate $\Lambda_{\text{LR}}(\mathbf{g})$ \cite{zhou2019approximating,li2024application} on raw measurement data. 
This will be accomplished by use
of an appropriately designed CNN-based classifier~\cite{zhou2019approximating}. Specifically, convolutional layers are gradually added to the classifier until the model's detection performance converges. The use of CNN-approximated IO (CNN-IO) has been successfully applied to raw measurement data in our previous study~\cite{li2024application}.

%\vspace{-0.3cm}
\section{Numerical Studies}
%\vspace{-0.1cm}
Computer-simulation studies were conducted to investigate the use of CNN-IOs to establish task-based performance bounds for image reconstruction based on binary signal detection tasks.
The study employed a stylized simulation of brain magnetic resonance imaging (MRI). 
Simulated $k$-space data were computed using a discrete-to-discrete (D-D) forward operator corresponding to a multi-coil sensitivity encoding (SENSE) parallel MRI system. Details regarding the simulated multi-coil MRI system, stochastic object models, and noise statistics are described below.
%SKE/BKE and BKS binary signal detection task were considered to validate the performance of the CNN-IO on raw measurement data.

%by comparing the test statistics from CNN-IO with those from analytical computation.}
%In addition, three BKS binary signal detection tasks with gradually increased task complexities (including both SKE and SKS) were employed to investigate the impact of dose reduction on estimated task-based performance bounds.

% \textcolor{blue}{A SKE/BKE binary signal detection task was considered to validate the performance of the CNN-IO on raw measurement data by comparing the test statistics from CNN-IO with those from analytical computation.}
% In addition, three BKS binary signal detection tasks with gradually increased task complexities (including both SKE and SKS) were employed to investigate the impact of dose reduction on estimated task-based performance bounds.

%\vspace{-0.2cm}
\subsection{Stylized multi-coil SENSE MRI systems}
\label{ssec:system}
Multi-coil SENSE MRI systems with 8 coils were modeled \cite{ohliger2006introduction}. For the $i^{th}$ coil, the corresponding \textit{k}-space measurement $\mathbf{g}_i$ was simulated as $\mathbf{g}_i=\mathbf{\Phi FS}_i \mathbf{f+n}_i$, where $\mathbf{f}$ was the to-be-imaged object, $\mathbf{\Phi}$ was the Cartesian sampling mask~\cite{zbontar2018fastmri}, $\mathbf{F}$ represented the discrete Fourier transform, $\mathbf{S}_i$ was the simulated coil sensitivity map~\cite{guerquin2011realistic}, and $\mathbf{n}_i$ was complex-valued Gaussian noise with standard deviation of 15. 

%\vspace{-0.2cm}
\subsection{Stochastic object model and signal}
\begin{figure}[h]
    \centering
    \includegraphics[width=0.8\textwidth]{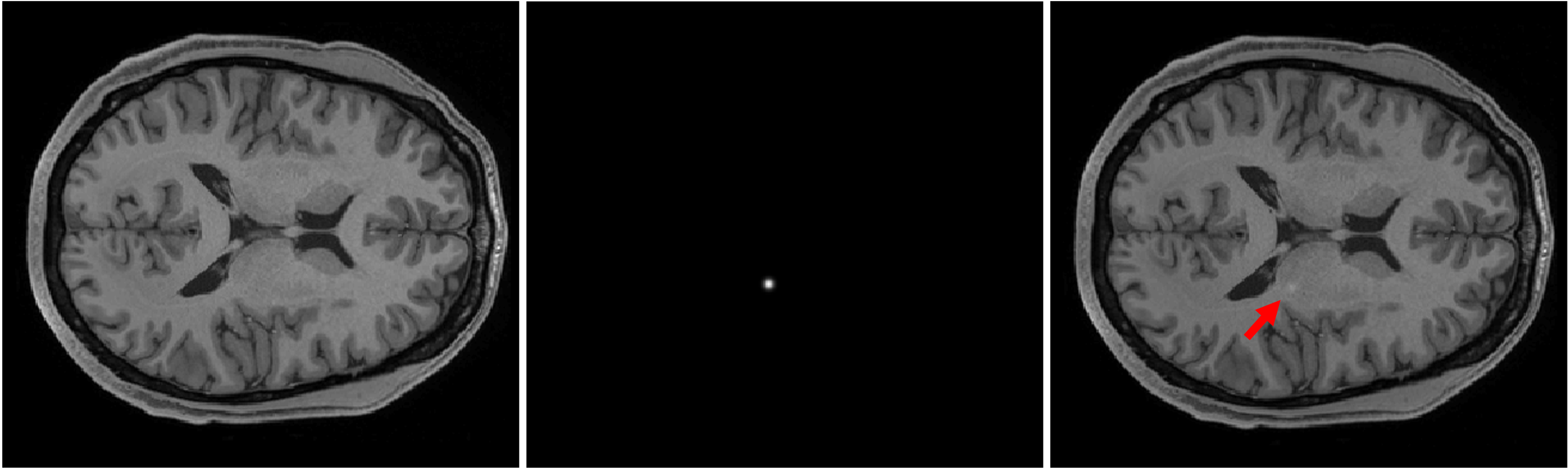}
        \caption{From left to right: a realization of background object; a realization of signal; a signal-present object created by combining the background and signal.
        The red arrow indicates the inserted signal.
                }
                %\vspace{-0.5cm}   
    \label{fig:sample}
\end{figure}

%\vspace{-0.1cm}
A stochastic object model (SOM) was employed to create ensembles of to-be-imaged objects. Specifically, the SOM described 2D axial slices of human brain MRI images. A diffusion model, i.e., denoising diffusion probabilistic model (DDPM)~\cite{ho2020denoising}, was employed to establish the SOM.  
It has been reported that compared with other generative models, the recently developed diffusion models can generate images with better visual quality \cite{ho2020denoising}.
The DDPM model was trained by use of the axial brain MRI slices from the Human Connectome Project (HCP) dataset.
In our study, only the slices that contain the cerebrospinal fluid (CSF) area were considered to minimize object variability due to slice location and instead emphasize differences between patients. 
Gaussian signals with random locations in the white matter area were considered in our study as the to-be-detected signal. The standard deviation of the Gaussian was 2 mm and its amplitude was 0.7.
Figure~\ref{fig:sample} shows a realization of background, signal, and signal-present objects. 

\subsection{Impact of acceleration factor}
\label{ssec:task}
% A SKE/BKE binary signal detection task was employed to validate the effectiveness of the CNN-IO on raw measurement data. 
% For the SKE/BKE task, a realization of the employed SOM and stochastic lesion were considered as the deterministic background and signal.
% The measurement noise was i.i.d. Gaussian with a standard deviation of 0.6. For this SKE/BKE task, the data space IO test statistic can be analytically computed on raw measurement data and was employed for comparison in this validation study.

A study was considered to investigate the impact of acceleration factors on the established task-based performance bounds. The acceleration factor was gradually increased from 1x (fully sampled) to 12x, and the corresponding impact on the established bounds was investigated.
Here, SKS/BKS binary signal detection tasks with random signal locations were considered in this study. Realizations of the established SOM were employed as the random background.
To assess the validity of the performance bounds for image reconstruction methods, both U-Net-based reconstruction methods \cite{jin2017deep} and root sum-of-square (rSOS) were considered as examples of to-be-evaluated reconstruction methods.

\section{Results}
%\subsection{Validate the CNN-IO acting on the raw tomographic measurement data}

% \begin{figure}[h]
% \centering
% \includegraphics[height=.4\linewidth]{figure/bke_auc.png}
%   \caption{ROC curves produced by the analytical computation (blue) and the data space CNN-IO (red) for the SKE/BKE task.}
%   \label{fig:result_bke}
% \end{figure}

% \begin{figure}
%     \centering
%     \includegraphics[width=0.9\textwidth]{figure/recon_physical.png}
%         \caption{Examples of FBP and U-Net reconstructed images. From left to right: Realizations of reconstructed images with 256, 128, 64, and 32 tomographic views.
%         %The SSIMs between the FBP-reconstruction and U-net based reconstruction to the object are 0.7879 and 0.9996.
%                 }
%                 %\vspace{-0.3cm}   
%     \label{fig:recon_sample}
% \end{figure}

% \begin{figure}[h]
% \centering
% \includegraphics[height=.4\linewidth]{figure/sparse_view_spie.png}
%   \caption{Relationships between AUC and four numbers of views when three BKS tasks were employed.}
%   \label{fig:result_dose}
% \end{figure}

\begin{figure}
\begin{floatrow}
\ffigbox{%
  \includegraphics[height=.6\linewidth]{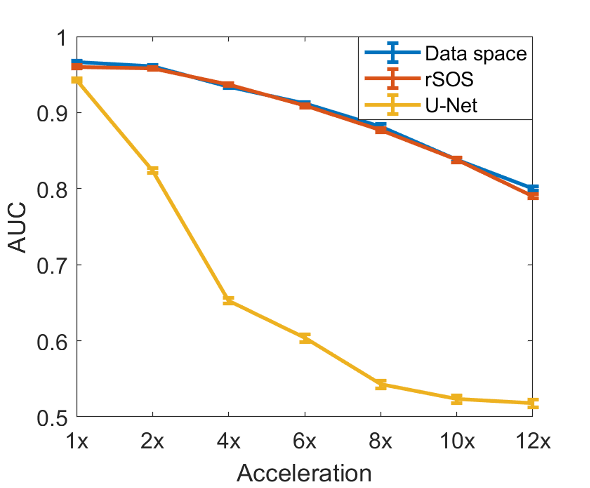}
}{%
  \caption{The impact of acceleration factors on task-based measures of IQ was investigated for the considered binary signal detection task.
Both CNN-IOs on raw measurement (blue), rSOS reconstructed images (red), and U-Net reconstructed images (yellow) were considered.}%
}
\label{fig:auc_acce}
\capbtabbox{%
   \begin{tabular}{c| l| c |c | c | c}
    \hline\hline
    \multicolumn{2}{c|}{Acceleration} & 1x & 4x & 8x & 12x\\
    \hline
    \multirow{2}{*}{RMSE} 
    &rSOS &0.0971 &0.1080 &0.1302 &0.1548 \\
      &U-Net &\textbf{0.0163} &\textbf{0.0230} &\textbf{0.0264} &\textbf{0.0318}\\
    \hline
    \multirow{2}{*}{SSIM} 
    &rSOS &0.7048 &0.6764 &0.6066 &0.5530\\
      &U-Net &\textbf{0.9833} &\textbf{0.9696} &\textbf{0.9647} &\textbf{0.9543}\\
	\hline\hline
	\end{tabular}
}{\vspace{1.5cm}
  \caption{The relationships between traditional measures (RMSE, SSIM) and acceleration factors were quantified. The U-Net-based methods greatly improved traditional IQ measures but not task-based IQ measures as compared to the rSOS images.}%
}
\end{floatrow}
\end{figure}

\begin{figure}
    \centering
    \includegraphics[width=0.9\textwidth]{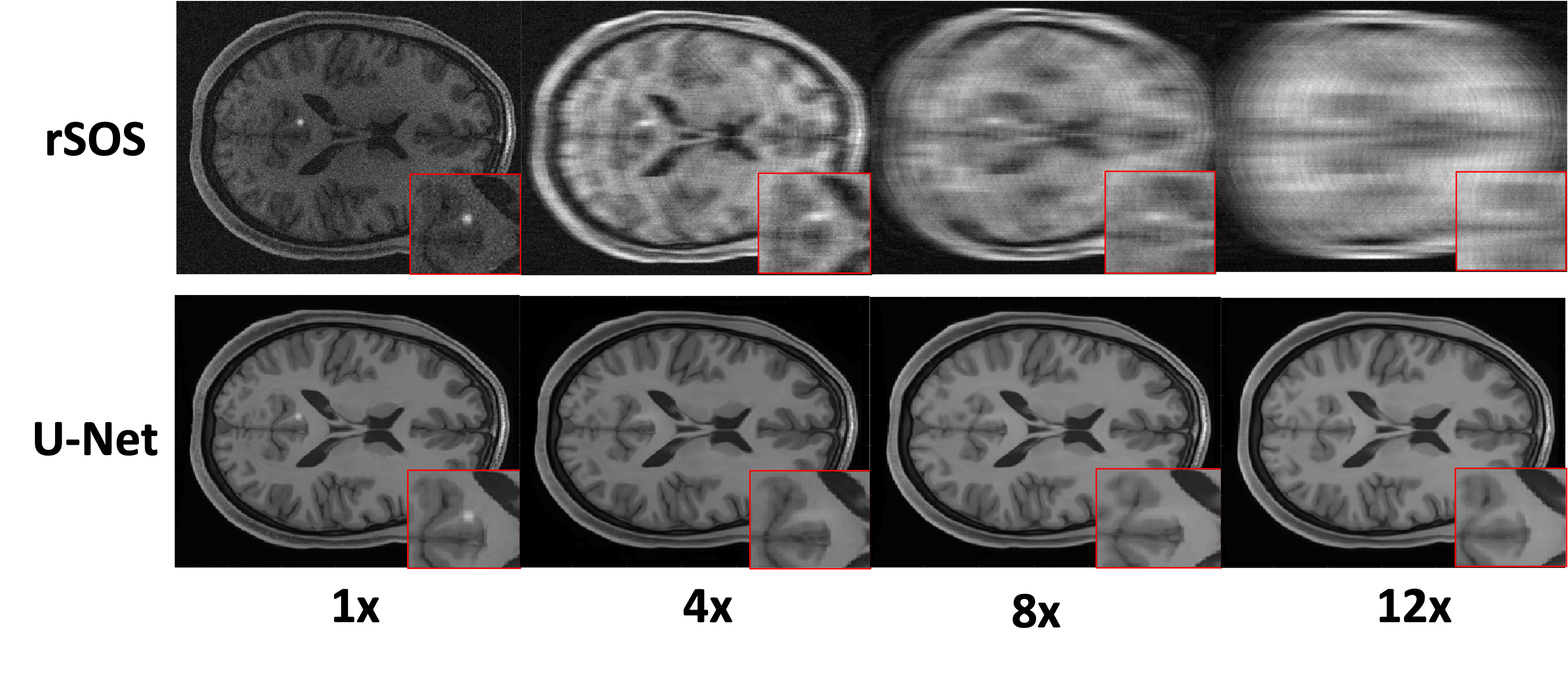}
    \vspace{-0.3cm}

        \caption{Examples of rSOS and U-Net reconstructed images. From left to right: Realizations of reconstructed images with 1x (fully sampled), 4x, 8x, and 12x acceleration factors. The U-net-based method greatly improved visual appearances as compared with the rSOS images, but the lesion signal disappeared after 4x or more accelerations.
        %The SSIMs between the FBP-reconstruction and U-net based reconstruction to the object are 0.7879 and 0.9996.
                }
                \vspace{-0.3cm}   
    \label{fig:recon_sample}
\end{figure}

%\subsection{Investigate the impact of dose reduction and task complexity on the estimated task-based performance bounds}

Figure \textcolor{blue}{2} shows the estimated task-based performance bounds for different acceleration factors (1x-12x).
As expected, for all cases, it was observed that the established bounds decreased as a function of the acceleration factor. 
As such, estimation of IO performance can provide valuable guidance when designing data-acquisition techniques and reconstruction methods for a certain task. 

In addition, as shown in Fig. \textcolor{blue}{2}, Tab.~\textcolor{blue}{1}, and Fig.~\ref{fig:recon_sample}, the U-Net-based method greatly improved traditional IQ measures and visual appearances but not task-based IQ measures when compared with the rSOS method for all considered acceleration factors.  This confirms the fact that traditional IQ measures may not correlate with specific task-based measures of IQ.

\section{NEW OR BREAKTHROUGH RESULTS TO BE PRESENTED}
This work explored the application of data space CNN-IOs in a multi-coil SENSE parallel MRI system to estimate performance bounds for DLIRMs in location-unknown signal detection tasks. Brain object variability was characterized by an advanced diffusion model.
This work will help advance the use of data space IO for analyses of imaging technologies under clinically relevant conditions.

\vspace{-0.2in}
\section{Conclusion}
In this paper, studies were designed to investigate the use of CNN-IO in raw data space to estimate task-based performance bounds for image reconstruction methods, under a realistic multi-coil SENSE MRI context.
Advanced diffusion models were employed to establish the SOM that generated realistic object variability for location-unknown signal detection tasks.
A preliminary numerical study demonstrated that the CNN-IO could be potentially employed to estimate the task-based performance bounds for various acceleration factors.
Importantly, our study demonstrated that although traditional IQ metrics such as RMSE and SSIM can reflect the effectiveness of DLIRMs relative to data fidelity and visual perception, they may not accurately reflect the effectiveness in preserving task-relevant information. In contrast, our results showed that the U-Net-based DLIRM could not restore small-sized signals as the acceleration factor increased to 4 or above. 
Therefore, the effectiveness of novel DLIRMs should not be determined solely using data fidelity or perpetual metrics, and task-based figures of merits must also be considered.
%This study demonstrates that the effectiveness of modern DLIRMs should be carefully assessed by use of task-based figures of metrics. 

\vspace{-0.3cm}

\section{Acknowledgments}
\begin{minipage}{\textwidth}
This work was supported in part by NIH Awards P41EB031772 (sub-project 6366), R01EB034249, R01CA233873, 
R01CA287778, and R56DE033344.
Kaiyan Li acknowledges funding by an appointment to the Research Participation Program at the Center for Devices and Radiological Health administered by the Oak Ridge Institute for Science and Education through an inter-agency agreement between the U.S. Department of Energy and U.S. Food and Drug Administration.
\end{minipage}
% \section{Acknowledgments }
% This work is original and has not been submitted for publication or presentation elsewhere. This work was supported in part by NIH awards R01EB020604, R01EB023045, R01NS102213, R01CA233873, R21CA223799, Cancer Center at Illinois seed grant, Jump ARCHES award, and DoD Award No. E01 W81XWH-21-1-0062.
% This work was supported in part by award NIH R01EB020604, R01EB023045, R01NS102213, R01CA233873, and R21CA223799.

% References
\footnotesize{
%\scriptsize{
\bibliography{denoising} % bibliography data in report.bib

\begin{thebibliography}{10}

\bibitem{barrett2013foundations}
Barrett, H.~H. and Myers, K.~J.,  [{\em Foundations of image science}{\nolinebreak\hspace{0.1em}]}, John Wiley \& Sons (2013).

\bibitem{kupinski2003ideal}
Kupinski, M.~A., Hoppin, J.~W., Clarkson, E., and Barrett, H.~H., ``Ideal-observer computation in medical imaging with use of markov-chain monte carlo techniques,'' {\em JOSA A}~{\bf 20}(3),  430--438 (2003).

\bibitem{li2021assessing}
Li, K., Zhou, W., Li, H., and Anastasio, M.~A., ``Assessing the impact of deep neural network-based image denoising on binary signal detection tasks,'' {\em IEEE Transactions on Medical Imaging}  (2021).

\bibitem{zhou2019approximating}
Zhou, W., Li, H., and Anastasio, M.~A., ``Approximating the ideal observer and hotelling observer for binary signal detection tasks by use of supervised learning methods,'' {\em IEEE transactions on medical imaging}~{\bf 38}(10),  2456--2468 (2019).

\bibitem{li2022task}
Li, K., Li, H., and Anastasio, M.~A., ``A task-informed model training method for deep neural network-based image denoising,'' in [{\em Medical Imaging 2022: Image Perception, Observer Performance, and Technology Assessment}{\nolinebreak\hspace{0.1em}]},   {\bf 12035},  249--255, SPIE (2022).

\bibitem{li2022impact}
Li, K., Li, H., and Anastasio, M.~A., ``On the impact of incorporating task-information in learning-based image denoising,'' {\em arXiv preprint arXiv:2211.13303}  (2022).

\bibitem{li2024application}
Li, K., Villa, U., Li, H., and Anastasio, M.~A., ``Application of learned ideal observers for estimating task-based performance bounds for computed imaging systems,'' {\em Journal of Medical Imaging}~{\bf 11}(2),  026002--026002 (2024).

\bibitem{ohliger2006introduction}
Ohliger, M.~A. and Sodickson, D.~K., ``An introduction to coil array design for parallel mri,'' {\em NMR in Biomedicine: An International Journal Devoted to the Development and Application of Magnetic Resonance In vivo}~{\bf 19}(3),  300--315 (2006).

\bibitem{zbontar2018fastmri}
Zbontar, J., Knoll, F., Sriram, A., Murrell, T., Huang, Z., Muckley, M.~J., Defazio, A., Stern, R., Johnson, P., Bruno, M., et~al., ``fastmri: An open dataset and benchmarks for accelerated mri,'' {\em arXiv preprint arXiv:1811.08839}  (2018).

\bibitem{guerquin2011realistic}
Guerquin-Kern, M., Lejeune, L., Pruessmann, K.~P., and Unser, M., ``Realistic analytical phantoms for parallel magnetic resonance imaging,'' {\em IEEE Transactions on Medical Imaging}~{\bf 31}(3),  626--636 (2011).

\bibitem{ho2020denoising}
Ho, J., Jain, A., and Abbeel, P., ``Denoising diffusion probabilistic models,'' {\em Advances in neural information processing systems}~{\bf 33},  6840--6851 (2020).

\bibitem{jin2017deep}
Jin, K.~H., McCann, M.~T., Froustey, E., and Unser, M., ``Deep convolutional neural network for inverse problems in imaging,'' {\em IEEE Transactions on Image Processing}~{\bf 26}(9),  4509--4522 (2017).

\end{thebibliography}
\bibliographystyle{spiebib} % makes bibtex use spiebib.bst
}

\end{document}